\documentclass{iau}
\usepackage{graphicx}
\usepackage{natbib}

\title[Feeding and feedback in LLAGNs] %% give here short title %%
{Feeding and Small-scale Feedback \\ in Low-Luminosity AGNs.}

\author[Roman V. Shcherbakov et al.]   %% give here short author list %%
{Roman V. Shcherbakov $^{1,2}$, Frederick K. Baganoff $^3$, Ka-Wah Wong $^4$ \and Jimmy Irwin $^4$}

\affiliation{$^1$ Department of Astronomy, University of Maryland, College Park, MD 20742-2421, USA \\ email: {\tt roman@astro.umd.edu} \\[\affilskip]
$^2$Hubble Fellow\\[\affilskip]
$^3$Kavli Institute for Astrophysics and Space Research, Massachusetts Institute of Technology, Cambridge, MA, 02139 \\[\affilskip]
$^4$Department of Physics and Astronomy, University of Alabama, Tuscaloosa, AL 35487, USA}

\pubyear{2012}
\volume{290}  %% insert here IAU Symposium No.
\jname{Feeding compact objects: Accretion on all scales}
\editors{C.M. Zhang, T. Belloni, M. M\'endez \& S.N. Zhang, eds.}

\begin{document}
\maketitle
\begin{abstract}
The unmatched X-ray resolution of \textit{Chandra} allows probing the gas flow near quiescent supermassive black holes (BHs).
The radius of BH gravitational influence on gas, called the Bondi radius, is resolved in Sgr A* and NGC 3115.
Shallow accretion flow density profiles $n\propto r^{-\beta}$ with $\beta=0.7-1.0$ were found for Sgr A* and NGC 3115 with the help of \textit{Chandra}.
We construct self-consistent models with gas feeding and dynamics from near the Bondi radius to the event horizon to explain the observations.
Gas is mainly supplied to the region by hot colliding stellar winds. Small-scale feedback such as conduction effectively flattens the density profile from steep $\beta=1.5$
in a Bondi flow. We further constrain density and temperature profiles using the observed radio/sub-mm radiation emitted near the event horizon.
We discuss the present state of our numerical model and its qualitative features, such as the role of the galactic gravitational potential
and the random motion of wind-emitting stars.
\end{abstract}

\firstsection
\section{Introduction}
As indicated by both the theory and the observations a typical active galactic nucleus (AGN) is not particularly active \citep{Ho:2008rev}.
 The median Eddington ratio in the Palomar survey of nearby AGNs is $\lambda=L_{\rm bol}/L_{\rm Edd}\sim10^{-5}$ \citep{Ho:2009se}. Palomar survey still misses AGNs like Sgr A*, which has an even lower ratio $\lambda\sim10^{-8}$ \citep{Narayan:2002ak}. Extensive radio observations help to identify weaker AGNs. While Sgr A* is an established radio emitter,
the activity in the nuclei of, e.g., M31 and NGC3115 is barely detected with the Jansky Very Large Array (\citealt{Crane:1992mj}; Wrobel \& Nyland, 2012, subm.). AGNs have a short duty cycle \citep{Greene:2007aa} and are in a low-luminosity state most of the time. AGNs enter a short bright quasar phase, when galaxy mergers feed the associated BHs with gas \citep{Hopkins:2006po,Hopkins:2008lh}. AGNs shut off quickly after the mergers due to the large-scale feedback \citep{Hopkins:2009kq}. The material in an isolated galaxy can accrete in various ways onto the central BH, which may power Seyfert galaxies (e.g., \citealt{Hopkins:2006mi}). When the gas supply from other sources is small, the BH can be fed mostly by stellar winds \citep{Hopkins:2006mi,Ho:2009se}. Typical mass loss rates from nuclear star clusters lead to very sub-Eddington accretion rates. Colliding stellar winds produce a hot tenuous medium, which might have a long cooling timescale. In the absence of cooling the accretion occurs as a radiatively inefficient accretion flow (RIAF) \citep{Narayan:1998re,Quataert:2001op}. Modeling BH feeding from stellar winds and the associated accretion flow in low-luminosity AGNs (LLAGNs) is the purpose of the present work. We discuss the features of the accretion flow, which delivers only a small fraction of available matter to the BH event horizon. We construct a semi-analytic radial model and apply it to Sgr A* and NGC3115.
\section{Supply and Accretion of Stellar Winds}
\begin{figure}[htbp]
    \centering\includegraphics{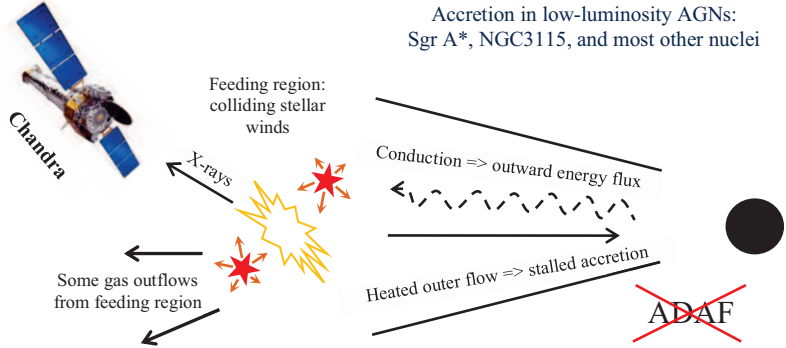}
    \caption{The scheme of accretion in LLAGNs. Colliding stellar winds provide matter into the feeding region. Some of that matter outflows from the region, while a fraction accretes.
    The heat from the inner flow leads to supervirial outer temperature and stalled accretion. Image of \textit{Chandra} satellite was produced by \textit{Chandra} team.}
    \label{fig:scheme}
\end{figure}
The matter is supplied near the radius of BH gravitational influence, the Bondi radius $r_B=GM_{BH}/c^2$, by stellar winds.
The mass loss rate $\dot{m}_{\rm ej}$ in winds can be estimated based on properties of the nuclear stellar population. The predicted $\dot{m}_{\rm ej}$ rate depends on the age of the stellar population \citep{Jungwiert:2001}, the initial mass function, metallicity, and the stellar evolution model \citep{Leitner:2011ad}. Approximate correlations exist of mass loss rate with V-band or B-band luminosity \citep{Ciotti_winds:1991,Padovani:1993de,Ho:2009se} for old stellar populations. Such relations are relatively easy to apply, since surface brightness in these bands was determined in many galactic nuclei as a by-product of searching for supermassive BHs \citep{Kormendy:1995}. In the case of Sgr A* the mass loss rate was directly computed from line properties of individual wind-emitting stars \citep{Martins:2007}.

As simulations show (e.g., \citealt{Cuadra_winds:2008}), the gas may exist in a quasi-steady state over long periods of time near the Bondi radius.
Accretion onto the BH and outflow balance the injection of stellar winds.  The gas energetics is determined by the velocity of colliding stellar winds, the gravitational potentials of the BH and the galaxy, and the energy flux from the inner accretion flow.  The wind-producing stars move in the gravitational potential at a speed $v_{st}$, while expelling winds at relative velocity $v_w$. Assuming no preferred direction of orbital angular momentum of stars, the specific energy of colliding winds is $<v_w^2+v_{st}^2>/2$, where $<>$ denotes the ensemble average. The gravitational potential of the galaxy is essential for setting the quasi-steady gas density. The galactic potential overtakes the BH potential at $r\sim20r_B$ for Sgr A* \citep{Schodel:2003jk} and at $r\sim2r_B$ for NGC3115 (Shcherbakov et al. 2012, in prep). Gravity creates a potential barrier for the outflow to escape from the feeding region. This potential barrier can be substantially underestimated, if the galactic potential is neglected. The percentage of the outflowing material may greatly vary. The accretion rate onto Sgr A* is $\sim 10^{-8}M_\odot{\rm yr}^{-1}$ \citep{Shcherbakov:2012appl}, while the cluster wind mass loss rate $10^{-3}M_\odot{\rm yr}^{-1}$ \citep{Martins:2007} is much higher. Lastly, the energy balance can be modified by the deposition of energy diffused from the inner accretion flow. Outward energy flux can flatten the density profile and shut off the accretion. Also it can help drive an outflow from the feeding region. \cite{Blandford:1999} state that "the binding energy of a gram of gas at a few $r_{\rm g}$ drives off $100$~kg of gas from $10^5$ $r_{\rm g}$", where $r_{\rm g}$ is a gravitational radius. The reason for such a dramatic influence can be understood from pressure balance. The balance of the gas pressure and the gravitational force reads
\begin{equation}
\frac{1}n\frac{\partial p}{\partial r}=\frac{\partial (n k_B T)}{n \partial r}=-\frac{G M m_p}{r^2},
\end{equation} where $n$ is particle density. The density profile is usually approximated as a power-law $n\propto r^{-\beta}$. As the temperature increases slightly due to the deposited energy, the density slope $\beta$ gets correspondingly shallower. A $25\%$ temperature increase causes $\beta$ to drop from $1.5$ to $1.0$, which leads to $300$ times lower accretion rate $\dot{M}$ for $r_B/r_{\rm g}=10^5$.

The origin of the energy flux may vary depending on gas density. Turbulent diffusion/convection is traditionally invoked to account for the energy flux \citep{Narayan:2000tr,Quataert:2000er}.
The flow is intrinsically turbulent and the large-scale eddies effectively transport the energy outwards.
Heat conduction takes over for turbulent diffusion at low gas density, when the mean free path of electrons becomes comparable to the flow size \citep{Sharma_spherical:2008}. Conductive heat flux is significant in turbulent magnetized flows, where the random magnetic field does not effectively trap the electrons \citep{Medvedev:2001,Ruskkowski:2011ak}.
As the problem is complex, and encompasses a huge range of radii $r_B/r_{\rm g}\sim10^5-10^6$ preventing the direct numerical simulations, we developed a semi-analytic radial model to quantify the accretion in LLAGNs.
\section{Inflow-outflow Model with Conduction and Applications}
The description of our first model can be found in \cite{Shcherbakov:2010cond}. We compute the radial profiles of gas density $\rho$, electron temperature $T_e$, and proton temperature $T_p$. We compile the radial profiles of mass and energy injection in winds. In the latest version of the model we include the random velocity of stars to compute the effective wind velocity. The gravitational potential is similarly improved to include the galactic potential. The electron temperature $T_e$ deviates down from $T_p$ in the inner flow at about $10^3r_{\rm g}$, where the electrons become relativistic. On the way to the BH different species are heated by $pdV$ work and by viscous energy dissipation. The flow had zero angular momentum in the first model, which may be a good assumption for the large values of effective viscosity $\alpha$ \citep{Shakura1973} or outside of the circularization radius. The transport of angular momentum and the associated heating are modeled self-consistently in the latest version. The energy is transported outwards by conduction. Conductive heat flux is proportional to the gradient of $T_e$, while conductivity at each radius $r$ is a small fraction of a product of that radius $r$ by the electron velocity $v_e$.

The hot accretion flow onto a BH has a specific radiation signature, which was observed in nearby LLAGNs.
The gas temperature in the feeding region is about $T\sim1$~keV \citep{Pellegrini:1995jk}, as produced by the colliding winds.
The hot gas radiates in X-rays via a variety of physical processes such as heavy ion line emission.
We use the emission model based on ATOMDB v2.0 \citep{Foster:2012we} to compute the spectrum in collisional equilibrium for an assumed metallicity $Z$. We employ custom-generated response matrices and photoelectric absorption to simulate the spectra as seen by \textit{Chandra} and to compare with the observed ones.
The X-ray spectrum is dominated by heavy ion emission for gas temperatures in $0.3-1$~keV range, which makes the estimate of gas density heavily dependent on metallicity. Inner density and temperature can be determined based on radiation from hot inner flow. The plasma emits mainly synchrotron radiation at the electron temperature $T_e\gtrsim 10^{10}$~K. Assuming that the radio emission at the highest observed frequency of radio/sub-mm bump comes from near the event horizon, we can estimate the electron temperature and density based on that frequency and correspondent flux. The slope between the inner and the outer flows is $\beta=0.80-0.90$ for Sgr A* \citep{Shcherbakov:2012appl} and $\beta=0.7-0.8$ for NGC3115 (Shcherbakov et al. 2012, in prep), while $\beta=1$ was found in the outer flow \citep{Wong:2011} of NGC3115.

\acknowledgements
NASA grant HST-HF-51298.01 (RVS) and Chandra grant GO1-12119X (FKB).

%\bibliographystyle{apj}
%\bibliography{BBL/refs_fin}

\begin{thebibliography}{31}
\expandafter\ifx\csname natexlab\endcsname\relax\def\natexlab#1{#1}\fi
\bibitem[{{Blandford} \& {Begelman}(1999)}]{Blandford:1999}{Blandford}, R.~D., \& {Begelman}, M.~C. 1999, \textit{MNRAS}, 303, L1
\bibitem[{{Ciotti} {et~al.}(1991){Ciotti}, {D'Ercole}, {Pellegrini}, \&  {Renzini}}]{Ciotti_winds:1991}{Ciotti}, L., {D'Ercole}, A., {Pellegrini}, S., \& {Renzini}, A. 1991, \textit{ApJ},  376, 380
\bibitem[{{Crane} {et~al.}(1992){Crane}, {Dickel}, \& {Cowan}}]{Crane:1992mj}{Crane}, P.~C., {Dickel}, J.~R., \& {Cowan}, J.~J. 1992, \textit{ApJ}, 390, L9
\bibitem[{{Cuadra} {et~al.}(2008){Cuadra}, {Nayakshin}, \&  {Martins}}]{Cuadra_winds:2008}{Cuadra}, J., {Nayakshin}, S., \& {Martins}, F. 2008, \textit{MNRAS}, 383, 458
\bibitem[{{Foster} {et~al.}(2012){Foster}, {Ji}, {Smith}, \&  {Brickhouse}}]{Foster:2012we}{Foster}, A.~R., {Ji}, L., {Smith}, R.~K., \& {Brickhouse}, N.~S. 2012, \textit{ApJ},  756, 128
\bibitem[{{Greene} \& {Ho}(2007)}]{Greene:2007aa}{Greene}, J.~E., \& {Ho}, L.~C. 2007, \textit{ApJ}, 667, 131
\bibitem[{{Ho}(2008)}]{Ho:2008rev}{Ho}, L.~C. 2008, \textit{Ann. Rev. Astron. Astr.}, 46, 475
\bibitem[{{Ho}(2009)}]{Ho:2009se}---. 2009, \textit{ApJ}, 699, 626
\bibitem[{{Hopkins} \& {Hernquist}(2006)}]{Hopkins:2006mi}{Hopkins}, P.~F., \& {Hernquist}, L. 2006, \textit{Ap. J. Supp.}, 166, 1
\bibitem[{{Hopkins} \& {Hernquist}(2009)}]{Hopkins:2009kq}---. 2009, \textit{ApJ}, 698, 1550
\bibitem[{{Hopkins} {et~al.}(2006){Hopkins}, {Hernquist}, {Cox}, {Di Matteo},  {Robertson}, \& {Springel}}]{Hopkins:2006po}
{Hopkins}, P.~F., {Hernquist}, L., {Cox}, T.~J., {Di Matteo}, T., {Robertson},  B., \& {Springel}, V. 2006, \textit{Ap. J. Supp.}, 163, 1
\bibitem[{{Hopkins} {et~al.}(2008){Hopkins}, {Hernquist}, {Cox}, \& {Kere{\v s}}}]{Hopkins:2008lh}
{Hopkins}, P.~F., {Hernquist}, L., {Cox}, T.~J., \& {Kere{\v s}}, D. 2008, \textit{Ap. J. Supp.}, 175, 356
\bibitem[{{Jungwiert} {et~al.}(2001){Jungwiert}, {Combes}, \& {Palou{\v  s}}}]{Jungwiert:2001}{Jungwiert}, B., {Combes}, F., \& {Palou{\v s}}, J. 2001, \textit{Astron. and  Astrophys.}, 376, 85
\bibitem[{{Kormendy} \& {Richstone}(1995)}]{Kormendy:1995}{Kormendy}, J., \& {Richstone}, D. 1995, \textit{Ann. Rev. Astron. Astr.}, 33, 581
\bibitem[{{Leitner} \& {Kravtsov}(2011)}]{Leitner:2011ad}{Leitner}, S.~N., \& {Kravtsov}, A.~V. 2011, \textit{ApJ}, 734, 48
\bibitem[{{Martins} {et~al.}(2007){Martins}, {Genzel}, {Hillier}, {Eisenhauer},  {Paumard}, {Gillessen}, {Ott}, \& {Trippe}}]{Martins:2007}
{Martins}, F., {Genzel}, R., {Hillier}, D.~J., {Eisenhauer}, F., {Paumard}, T.,  {Gillessen}, S., {Ott}, T., \& {Trippe}, S. 2007, \textit{Astron. and  Astrophys.},  468, 233
\bibitem[{{Narayan}(2002)}]{Narayan:2002ak}{Narayan}, R. 2002, in \textit{Lighthouses of the Universe: The Most Luminous Celestial  Objects and Their Use for Cosmology}, ed. M.~{Gilfanov}, R.~{Sunyeav}, \&  E.~{Churazov}, 405
\bibitem[{{Narayan} {et~al.}(2000){Narayan}, {Igumenshchev}, \&  {Abramowicz}}]{Narayan:2000tr}{Narayan}, R., {Igumenshchev}, I.~V., \& {Abramowicz}, M.~A. 2000, \textit{ApJ}, 539,  798
\bibitem[{{Narayan} {et~al.}(1998){Narayan}, {Mahadevan}, \&  {Quataert}}]{Narayan:1998re}{Narayan}, R., {Mahadevan}, R., \& {Quataert}, E. 1998, in \textit{Theory of Black Hole
  Accretion Disks}, ed. M.~A. {Abramowicz}, G.~{Bjornsson}, \& J.~E. {Pringle},  148
\bibitem[{{Narayan} \& {Medvedev}(2001)}]{Medvedev:2001}{Narayan}, R., \& {Medvedev}, M.~V. 2001, \textit{ApJ}, 562, L129
\bibitem[{{Padovani} \& {Matteucci}(1993)}]{Padovani:1993de}{Padovani}, P., \& {Matteucci}, F. 1993, \textit{ApJ}, 416, 26
\bibitem[{{Pellegrini}(2005)}]{Pellegrini:1995jk}{Pellegrini}, S. 2005, \textit{ApJ}, 624, 155
\bibitem[{{Quataert}(2001)}]{Quataert:2001op}{Quataert}, E. 2001, in \textit{Astronomical Society of the Pacific Conference Series},  Vol. 224, \textit{Probing the Physics of Active Galactic Nuclei}, ed. B. {Peterson}, R. {Pogge}, \& R. {Polidan}, 71
\bibitem[{{Quataert} \& {Gruzinov}(2000)}]{Quataert:2000er}{Quataert}, E., \& {Gruzinov}, A. 2000, \textit{ApJ}, 539, 809
\bibitem[{{Ruszkowski} \& {Oh}(2011)}]{Ruskkowski:2011ak}{Ruszkowski}, M., \& {Oh}, S.~P. 2011, \textit{MNRAS}, 414, 1493
\bibitem[{{Sch{\"o}del} {et~al.}(2003){Sch{\"o}del}, {Ott}, {Genzel}, {Eckart},  {Mouawad}, \& {Alexander}}]{Schodel:2003jk}{Sch{\"o}del}, R., {Ott}, T., {Genzel}, R., {Eckart}, A., {Mouawad}, N., \&  {Alexander}, T. 2003, \textit{ApJ}, 596, 1015
\bibitem[{Shakura \& Sunyaev(1973)}]{Shakura1973}Shakura, N.~I., \& Sunyaev, R.~A. 1973, \textit{Astron. and Astrophys.}, 24, 337
\bibitem[{{Sharma} {et~al.}(2008){Sharma}, {Quataert}, \&  {Stone}}]{Sharma_spherical:2008}{Sharma}, P., {Quataert}, E., \& {Stone}, J.~M. 2008, \textit{MNRAS}, 389, 1815
\bibitem[{{Shcherbakov} \& {Baganoff}(2010)}]{Shcherbakov:2010cond}{Shcherbakov}, R.~V., \& {Baganoff}, F.~K. 2010, \textit{ApJ}, 716, 504
\bibitem[{{Shcherbakov} {et~al.}(2012){Shcherbakov}, {Penna}, \&  {McKinney}}]{Shcherbakov:2012appl}{Shcherbakov}, R.~V., {Penna}, R.~F., \& {McKinney}, J.~C. 2012, \textit{ApJ}, 755, 133
\bibitem[Wong et al.(2011)]{Wong:2011} Wong, Ka-Wah, Irwin, Jimmy A., Yukita, Mihoko, Million, Evan T., Mathews, William G., \& Bregman, Joel N. 2011, \textit{ApJ}, 736, 23
\end{thebibliography}

\end{document}